\documentclass[aps,prf,twocolumn,a4paper,10pt,notitlepage,footinbib,superscriptaddress,longbibliography]{revtex4-1}
\usepackage[english]{babel}
\usepackage{lmodern}
\usepackage[latin9]{inputenc}
\usepackage{endnotes}
\usepackage{amssymb,amsmath,amsfonts}
\usepackage{textcomp}
\usepackage{braket}

\usepackage{graphicx}
\usepackage{dcolumn}
\usepackage{bm}
\usepackage{xcolor}

\begin{document}

\title{Concentrated phase emulsion with multi-core morphology under shear: A numerical study}

\author{A. Tiribocchi}
\affiliation{Center for Life Nano Science@La Sapienza, Istituto Italiano di Tecnologia, 00161 Roma, Italy}
\affiliation{Istituto per le Applicazioni del Calcolo CNR, via dei Taurini 19, Rome, Italy}
\author{A. Montessori}
\affiliation{Istituto per le Applicazioni del Calcolo CNR, via dei Taurini 19, Rome, Italy}
\author{F. Bonaccorso}
\affiliation{Center for Life Nano Science@La Sapienza, Istituto Italiano di Tecnologia, 00161 Roma, Italy}
\affiliation{Istituto per le Applicazioni del Calcolo CNR, via dei Taurini 19, Rome, Italy}
\affiliation{Department of Physics and INFN, University of Rome Tor Vergata, Via della Ricerca Scientifica 1, 00133, Rome, Italy}
\author{M. Lauricella}
\affiliation{Istituto per le Applicazioni del Calcolo CNR, via dei Taurini 19, Rome, Italy}
\author{S. Succi}
\affiliation{Center for Life Nano Science@La Sapienza, Istituto Italiano di Tecnologia, 00161 Roma, Italy}
\affiliation{Istituto per le Applicazioni del Calcolo CNR, via dei Taurini 19, Rome, Italy}
\affiliation{Institute for Applied Computational Science, John A. Paulson School of Engineering and Applied Sciences, Harvard University, Cambridge, USA}

\begin{abstract}

We numerically study the dynamic behavior under a symmetric shear flow of selected examples of concentrated phase emulsions with multi-core morphology confined within a microfluidic channel. A variety of new nonequilibrium steady states is reported. Under low shear rates, the emulsion is found to exhibit a solid-like behavior, in which cores display a periodic planetary-like motion with approximately equal angular velocity. At higher shear rates two steady states emerge, one in which  all inner cores align along the flow and become essentially motionless and a further one in which some cores accumulate near the outer interface and produce a dynamical elliptical-shaped ring chain, reminiscent of a treadmilling-like structure,  while others occupy the center of the emulsion. A quantitative description in terms of i) motion of the cores, ii) rate of deformation of the emulsion and iii) structure of the fluid flow within the channel is also provided.

\end{abstract}

\maketitle

\section{Introduction}

A multiple emulsion is an example of hierarchical soft fluid, consisting of smaller drops (often named as cores) dispersed within a larger one and stabilized by means of surfactants adsorbed onto the interfaces \cite{utada_2005,datta2014,vladi2017,anton2019}. A typical example in point is a collection of distinct water drops (either monodisperse or polidisperse) encapsulated within an oil one.

Recent microfluidic experiments have been capable of designing highly regular and well-defined multi-core emulsions with an extremely controlled procedure \cite{utada_2005,chu_2007,weitz_2009,weitz_2011,clegg_2016,vladi2017,anton2019}. The process essentially consists of two emulsification steps, one in which water droplets are dispersed in oil phase and a second one in which they are embedded in the same oil phase \cite{utada_2005,lee2016,vladi2017}. These results have paved the way to the use of multiple emulsions in a wide range of applications, ranging from pharmaceutics and food science, for delivery and controlled release of compounds \cite{pays2002,sela2009,mcall2013} and for the encapsulation of flavours \cite{garti2002,mclement2012,muschi2017}, to tissue engineering and cosmetics, for the realization of soft materials with high degree of porosity \cite{chung2012,costantini2014} and for the production of personal care items \cite{lee2004,datta2014}.  

Understanding the response of these systems to the effect of external fields, even under controlled experimental conditions, remains a crucial requirement for a purposeful design of such emulsion-based devices and for their correct functioning. This is particularly important in concentrated phase emulsions (CPEs) with multi-core morphology, in which, unlike the diluted regime, large portions of the material are occupied by fluid interfaces and the volume fraction of the cores is generally higher than $0.3$, thus long-range hydrodynamic interactions as well as droplet collisions cannot be easily neglected \cite{costantini2014,marmottant2009,montessori2019}. Indeed an external field, such as a shear flow, can produce relevant interface deformations  that propagate to the scale of the droplet and ultimately compromise structural integrity and mechanical stability of the whole emulsion.

From a theoretical standpoint, such physics can be almost exclusively investigated by means of numerical simulations built on specific computational models, due to the complicated structure of the equations governing the multiscale dynamics, typically ranging from the size of the interface to that of the microfluidic device \cite{succirev}. In contrast with the enormous progress achieved in the manufacturing of multiple emulsion \cite{utada_2005,garstecki2005,chu_2007,weitz2012,guzowski2015,ma2015,vladi2017}, theoretical studies about their red dynamic response under shear have been carried on only more recently (especially for double emulsions \cite{chen2013,wang2013,chen2015,kusu2020}) and have very partially covered the physics of concentrated phase emulsions \cite{pal2007,pal2011,tao2013}.

It is well-known, for example, that when a single-core emulsion is subject to a low/moderate shear flow, the external drop stretches and attains an elliptical shape at the steady state, while the internal core preserves reasonably well its spherical shape \cite{chen2013,chen2015}. Higher shear rates can lead to the breakup of the enveloping shell and the release of the internal drop \cite{smith2004}. When two distinct inner cores are included, an external shear flow can cause the formation of a fluid recirculation which triggers a periodic planetary-like motion of the cores, lasting over long periods of time  within the external droplet \cite{tiribocchi_pof}. Such features have been captured, with a good level of accuracy, by means of a careful combination of continuum models and numerical simulations, such as buondary integral methods \cite{wang2013} and lattice Boltzmann approaches \cite{tiribocchi_pof,kusu2020}. Yet, much less is known about the dynamics of a multiple emulsion under shear when the numer of inner cores augments and their concentration is far from the diluted regime.

In this paper we go one step further and numerically investigate, by using lattice Boltzmann simulations, the dynamic response of two-dimensional CPEs with multi-core morphology in which the area fraction occupied by the cores is higher than $30\%$. As in previous works \cite{marenduzzo_prl,tiribocchi_pof}, the physics is described by using a multiphase field model for an immiscible fluid mixture, in which a Landau free-energy is employed to calculate the thermodynamic forces (chemical potential and pressure tensor) governing the relaxation dynamics of the system.

By varying shear rate and number of cores, we provide evidence of new non-equilibrium steady states previously undetected. At low shear rates, internal cores exhibit a periodic motion, persistent over extended periods of time and triggered by the fluid vorticity. When the number of cores increases, a solid-like dynamic behavior can be envisaged, since all cores moves coherently with approximately equal angular speed. At intermediate shear rates, two different steady states are found. One in which cores form a motionless open chain aligned along the flow, and a further one, observed when their number is sufficiently high, in which some  aggregate near the external interface and arrange as an elliptical-shaped ring, moving as a treadmill, while others take place in the center of the emulsion and dynamically interact with the former ones. Further increasing the shear rate heavily squeezes the emulsion and forces the cores to arrange in a two-row structure displaying, once again, a periodic motion along a highly stretched elliptical path. This behavior also leads to significant modifications of the external interface, which, at the steady state, exhibits regular bulges due to the concurrent effect of hydrodynamic interactions and collisions with cores.  Such physics is characterized in terms of i) motion of the cores (time evolution of center of mass and its velocity), ii) rate of deformation of the cores and iii) structure of the fluid velocity.

The paper is structured as follows. In section \ref{II} we describe the computational model and the numerical setup used in the simulations, and in Section \ref{III} we discuss the results. We start by investigating the response under shear flow of a four-core emulsion, and present some selected non-equilibrium steady states observed at different values of shear rate. Then we report the results about non-equilibrium steady states observed in a higher complex emulsion, in which the number of cores is much increased. Some final remarks conclude the manuscript.

\section{The model}\label{II}

Here we outline the hydrodynamic model of a multi-core emulsion, a system  made of a collection of immiscible fluid droplets embedded in larger one. Its physics is captured by the following coarse-grain fields. A set of scalar phase-fields $\phi_i({\bf r},t)$, $i=1,....,N$ (where $N+1$ is the total number of drops and $N$ is the number of cores) represents the droplet density, positive within each drop and zero outside, while the average fluid velocity of both drops and solvent is described by a vector field ${\bf v}({\bf r},t)$ \cite{marenduzzo_prl,marenduzzo_soft,yeomans_prl,tiribocchi_pof}.

\subsection{Equations of motion}

The dynamics of the fields $\phi_i({\bf r},t)$ is governed by a set of convection-diffusion equations
\begin{equation}\label{CH_eqn}
\partial_t\phi_i+{\bf v}\cdot\nabla\phi_i=M\nabla^2\frac{\delta{\cal F}}{\delta\phi_i}
\end{equation}
where $M$ is the mobility and ${\cal F}=\int_VfdV$ is the total free energy describing the equilibrium properties of the fluid suspension.
A typical expression of the free energy density $f$ is given by \cite{degroot,lebon} 
\begin{equation}\label{freeE}
f= \frac{a}{4}\sum_i^N\phi_i^2(\phi_i-\phi_0)^2+\frac{k}{2}\sum_i^N(\nabla\phi_i)^2+\epsilon\sum_{i,j,i<j}\phi_i\phi_j,
\end{equation}
where the first term guarantees the existence of two coexisting minima, $\phi_i=\phi_0$ inside the $i$th droplet and $\phi_i=0$ outside,
and the second term stabilizes the droplet interface. The parameters $a$ and $k$ are two positive constants controlling surface tension and width of interface,
which are given by $\sigma=\sqrt{8ak/9}$ and $\xi=2\sqrt{k/2a}$ respectively \cite{widom,kruger}. Finally, 
the last term in Eq.(\ref{freeE}) is a soft-core repulsion contribution which penalizes the overlap of droplets, and whose strength
is gauged by the positive constant $\epsilon$.

The fluid velocity ${\bf v}({\bf r},t)$ obeys the continuity equation (in the incompressible limit)
\begin{equation}\label{CNT_eqn}
\nabla\cdot{\bf v}=0,
\end{equation}
and the Navier-Stokes equation
\begin{equation}\label{NAV_eqn}
  \rho\left(\frac{\partial}{\partial t}+{\bf v}\cdot\nabla\right){\bf v}=-\nabla p-\sum_i\phi_i\nabla\mu_i+\eta\nabla^2{\bf v}
\end{equation}
where $\rho$ is the fluid density, $p$ is its pressure, $\eta$ is the  viscosity of the fluid and $\mu_i\equiv\frac{\delta{\cal F}}{\delta\phi_i}$ is the chemical potential.

\subsection{Numerical details}

Eqs.~(\ref{CH_eqn}), (\ref{CNT_eqn}) and (\ref{NAV_eqn}) are solved by using a hybrid lattice Boltzmann (LB) method, in which the convection-diffusion equations are integrated by using
a finite difference approach while the continuity and the Navier-Stokes equations via a standard LB algorithm \cite{succi1,succi2,tiribocchi_epje,tiribocchi_pre2,tiribocchi_soft,montessori2015}.

\begin{figure}[htbp]
\includegraphics[width=1.\linewidth]{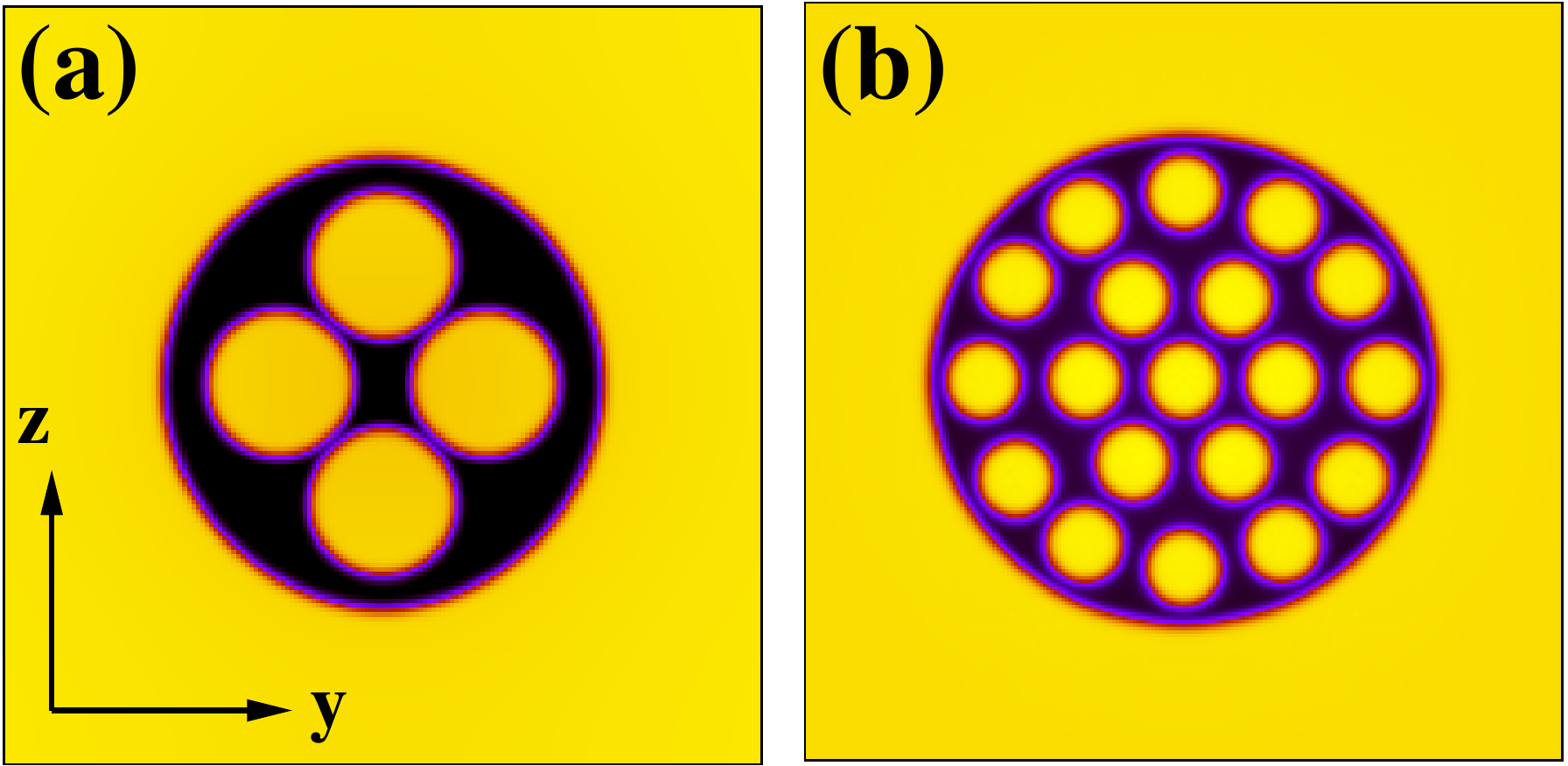}
\caption{Equilibrium configurations of a four-core emulsion (a) $N=4$, and a nineteen-core one (b) $N=19$. Only a section of the channel is shown. Colors correspond to the values of $\phi$, ranging from $0$ (black, middle phase) to $\simeq 2$ (yellow, inner and outer phase), while red lines indicate the droplet interface.}
\label{fig1} 
\end{figure}

Multi-core emulsions are sandwitched between two parallel flat walls placed at distance $L_z$ (see Fig.\ref{fig1}), where we set neutral wetting for $\phi_i$ and no-slip conditions for ${\bf v}$. The former one is achieved by imposing that ${\bf n}\cdot\nabla\mu_i|_{z=0,z=L_z}=0$ (no mass flux through the walls) and $\nabla(\nabla^2\phi_i)|_{z=0,z=L_z}=0$ (interface droplets perpendicular to the walls), where ${\bf n}$ is a unit vector normal to the boundaries pointing inward. The latter one means that $v_z(z=0,z=L_z)=0$.

In Fig.\ref{fig1} we show two examples of CPEs, inspired by experiments realized in the lab \cite{weitz2012,manovic2017,vladi2017,anton2019} and  made of approximately monodisperse cores arranged in a highly symmetric fashion. In (a) four cores (yellow), each of diameter $D_i=30$ lattice sites, are accommodated within a larger circular region (the second fluid component in black), of diameter $D_O=100$ lattice sites, in turn surrounded by an external fluid (yellow). Each field $\phi_i$ is positive and equal to $\simeq 2$ within each core and zero elsewhere, while the external fluid is positive out of the emulsion and zero within. In (b) a larger number of cores ($N=19$), each of diameter $D_i=20$ lattice sites, is included within a circular drop, of diameter $D_O=136$ lattice sites, immersed in a further fluid. In both cases, the area fraction $A_f=N\pi R_i^2/\pi R_O^2$ occupied by the cores is approximately $0.4$, considerably below the close packing limit ($\sim 0.74$), but sufficiently high to be far from the diluted regime and well within the concentrated one \cite{McClements_2019}. 

Such emulsions are initially relaxed towards a near-equilibrium state, and afterwards a symmetric shear is applied, by moving the top wall along the positive $y$-axis with velocity $v_w$ and the bottom one along the opposite direction with velocity $-v_w$. The resulting shear rate is $\dot{\gamma}=2v_w/L_z$. In our simulations $v_w$ ranges between $10^{-2}$ (low shear) to $4\times 10^{-2}$ (moderate/high shear), which means that, in a channel of transversal size $L_z=170$, $\dot{\gamma}$ varies from $\simeq 10^{-4}$ to $\simeq 3.5\times 10^{-4}$. The dynamic evolution of the emulsion is computed with respect to a dimensionless time $t^*=\dot{\gamma}(t-t_{eq})$, where $t_{eq}$ is the relaxation time after which the shear is imposed, approximately equal to $t_{eq}\simeq 10^5$ time-steps \cite{chen_2013,chen_2015,tiribocchi_pof}. If not elsewhere stated, the thermodynamic parameters used in our simulations are the following: $a=0.07$, $M=0.1$, $\eta=1.67$, $k=0.1$, $\epsilon=0.05$, $\Delta x=1$, $\Delta t=1$, where $\Delta x$ and $\Delta t$ are the lattice spacing and the integration time step. Dimensions of the lattice are set as follows: $L_y=400$ and $L_z=170$ for the four-core emulsion, while $L_y=800$ and $L_z=250$ for the nineteen-core emulsion.

Our system can be approximately mapped onto a multi-core emulsion in which internal drops, of diameter  $\simeq 10-50 \mu$m and surface tension ranging between $1-10$mN/m, are immersed in a fluid of viscosity $\simeq 10^{-1}$ Pa$\cdot$s, assumed equal for both continuous (background fluid) and dispersed phase (cores) \cite{marenduzzo_prl,utada_2005,guido_2010}. Typical speeds vary between $0.1-0.5$ mm/s under a shear rate of $0.1-1/s$. Finally, both capillary and Reynolds numbers vary between $0.1$ and $5$. If, for instance, $v=0.01$ (in simulation units), one gets $Ca=\frac{v\eta}{\sigma}\simeq 0.2$ and $Re=\frac{\rho vD_O}{\eta}\simeq 0.6 - 1.2$, assuming the diameter of the emulsion $D_O$ as a characteristic length, typically ranging from $50$ to $100$ lattice units.

\section{Results}\label{III}

We start by investigating the dynamic response under a symmetric shear flow of the concentrated phase emulsion shown in Fig.\ref{fig1}a and afterwards we discuss the case shown in Fig.\ref{fig1}b.

\subsection{Four-core emulsion}

In a previous work \cite{tiribocchi_pof} we have studied the dynamic behavior of a 2D multi-core emulsion in which two or three cores are dispersed within a larger drop at a relatively low area fraction $A_f$, ranging from $0.1$ to $0.2$. We have shown that, if the shear rate is kept sufficiently low to avoid the emulsion breakup, at the steady state the cores exhibit a periodic planetary-like motion, caused by a fluid vortex formed within the emulsion and triggered by the sheared structure of the flow. Here we show that such description holds only partially when the area fraction of the cores increases up to $0.4$ (like in a concentrated phase emulsion), since significant modifications of both emulsion shape and dynamics occur. 

\begin{figure}
\includegraphics[width=1.\linewidth]{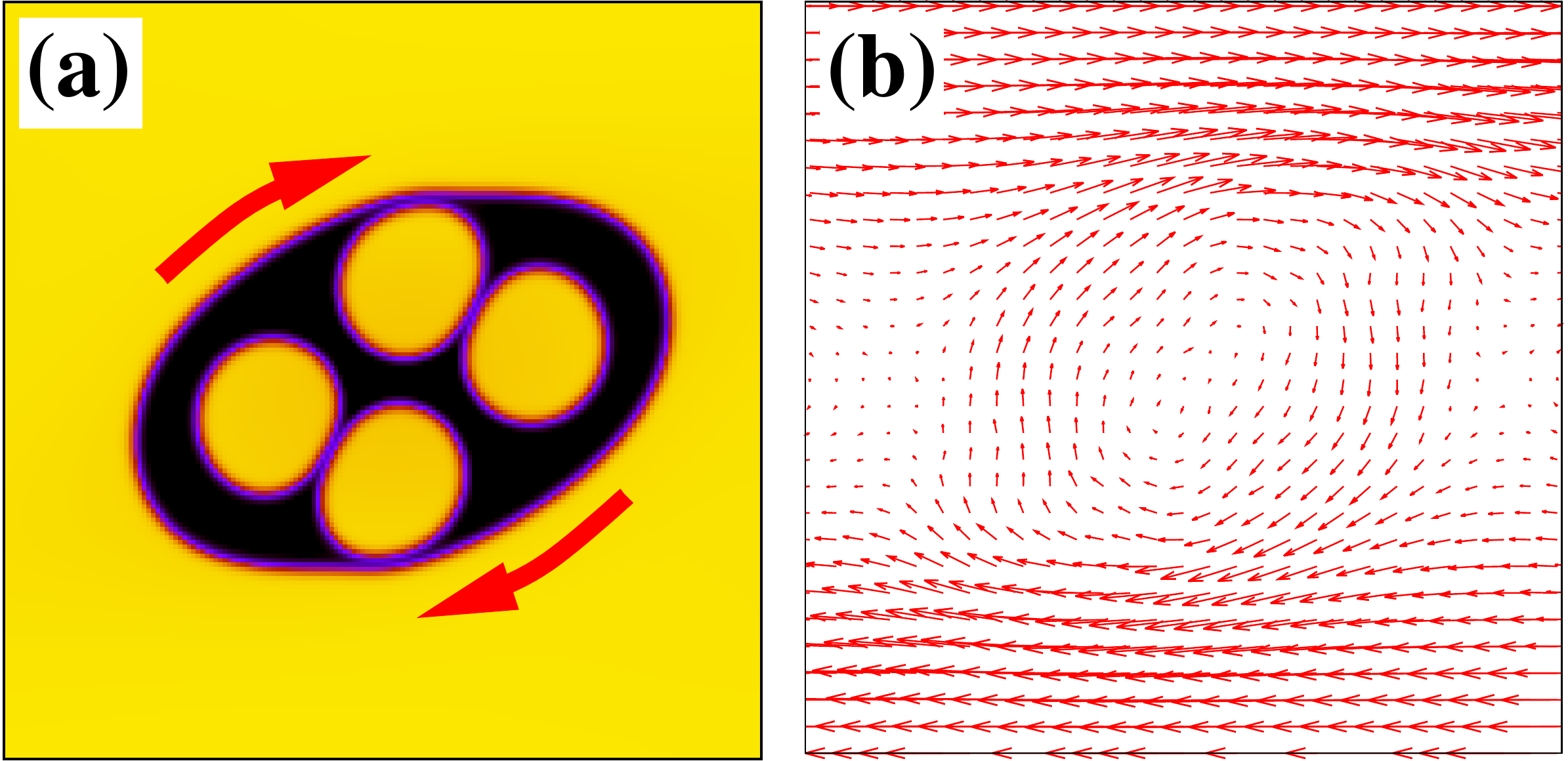}
\caption{(a) Typical nonequilibrium steady-state  of a four-core emulsion under a symmetric shear flow with $\dot{\gamma}\simeq 10^{-4}$ ($v_w=0.01$). A clockwise rotation of the internal cores is triggered by a fluid recirculation produced by the shear within the emulsion. Red arrows indicate the direction of rotation. (b) Steady state velocity field under shear. Snapshot is taken at $t^*=72$. Here $Ca\simeq 0.2$ and $Re\simeq 1.2$.}
 \label{fig2}
\end{figure}

\begin{figure}
\includegraphics[width=1.\linewidth]{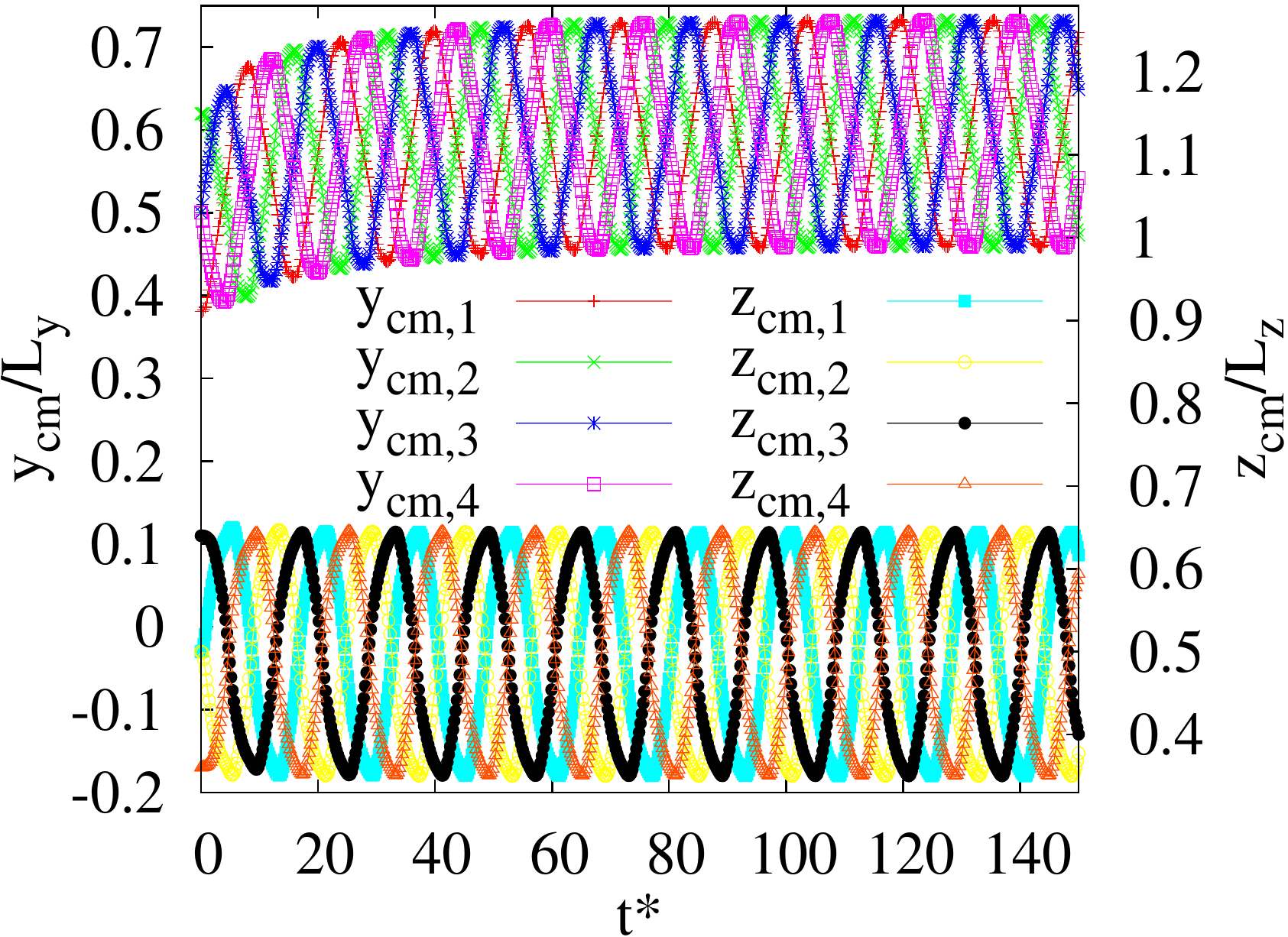}
\caption{Time evolution of the $y$ (left axis, top plots) and $z$ (right axis, bottom plots) components of the center of mass of the cores at the steady state. They are computed as $y_{cm,i}(t)=\frac{\sum_yy(t)\phi_i(y,z,t)}{\sum_y\phi_i(y,z,t)}$, and $z_{cm,i}(t)=\frac{\sum_zz(t)\phi_i(y,z,t)}{\sum_z\phi_i(y,z,t)}$, taken where $\phi_i>0.01$.}
 \label{fig3}
\end{figure}

In Fig.\ref{fig2} (and \cite{Suppl}, movie M1, where the full dynamics is reported) we show a typical steady state configuration of a four-core emulsion for $\dot{\gamma}\simeq 10^{-4}$. The motion proceeds essentially as discussed for emulsions with a lower number of cores. Once the shear is imposed, they acquire motion and periodically rotate clockwise in a self-repeating manner around a common center of mass, following approximately circular orbits within the surrounding fluid capsule, which acts as an effective confining bag. The time evolution of their center of mass (Fig.\ref{fig3}) shows that such dynamics is persistent, and no appreciable deviations from it are observed at late-time.

As long as the droplet shape remains sufficiently well defined (like a circle or an ellipse), shape deformations can be reasonably captured by the Taylor parameter $D=\frac{a-b}{a+b}$, where $a$ and $b$ are the length of the major and the minor axis. In Fig.\ref{fig3} we show the steady-state time evolution of $D$ of the external droplet and of one of the cores. The former is approximately four times larger than the latter (whose deformation remains negligible), and both exhibit time oscillations due to the reciprocal and recurring interactions among cores and with the external interface. This result proves, once again, that the outer droplet acts as an effective shield, preventing relevant deformations of the inner cores \cite{tiribocchi_pof}.  

\begin{figure}
\includegraphics[width=1.\linewidth]{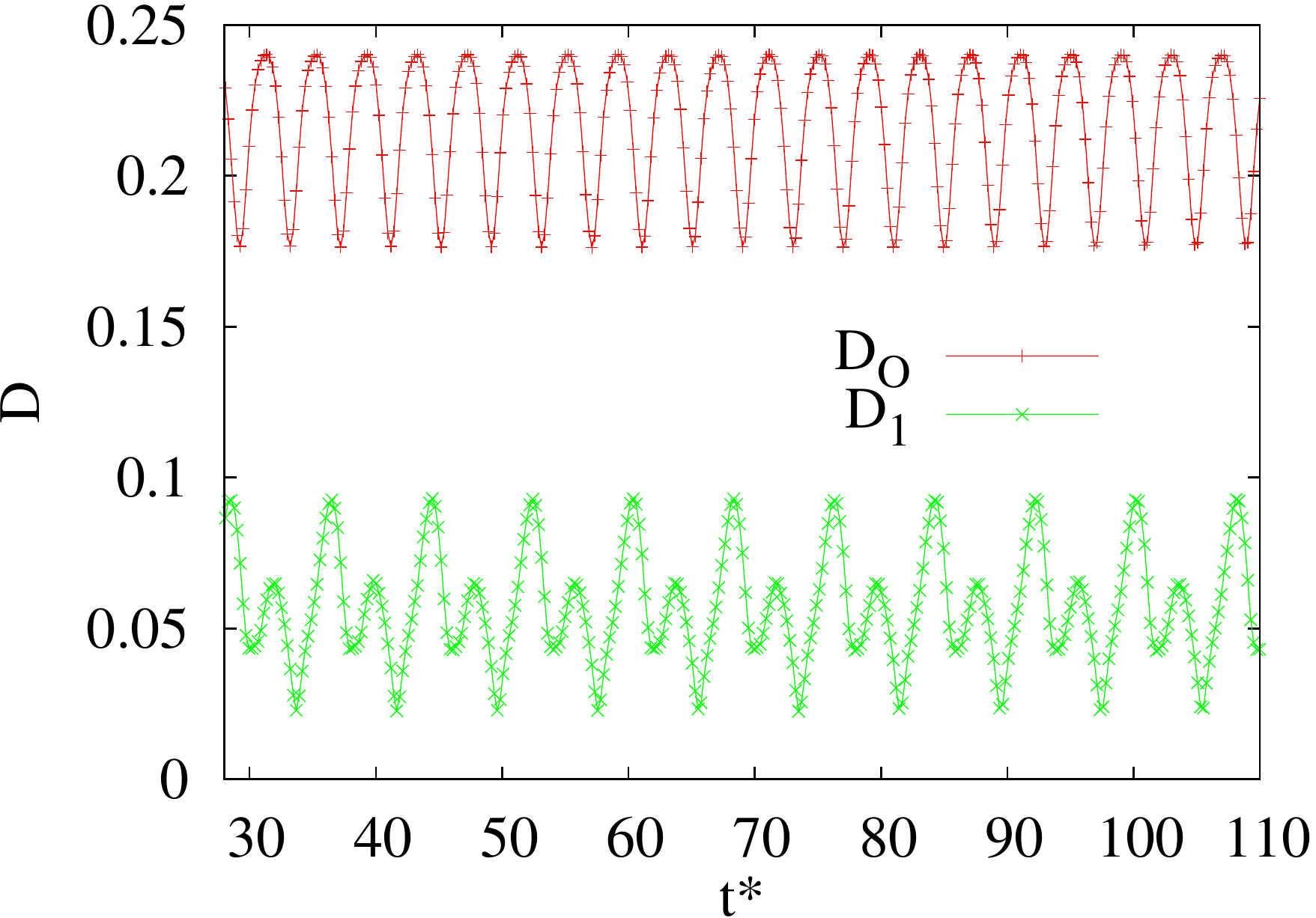}
\caption{Time evolution of the Taylor parameter of the outer droplet $D_O$ and of one of the cores $D_1$. Time oscillations are due to reciprocal interactions among internal droplets and with the external interface \cite{tiribocchi_pof}.}
 \label{fig4}
\end{figure}

However, how ``long-lived'' are such steady states? In other words, can the periodic dynamics of the cores turn into an alternative steady configuration before the inevitable rupture?

\begin{figure*}
\includegraphics[width=1.\linewidth]{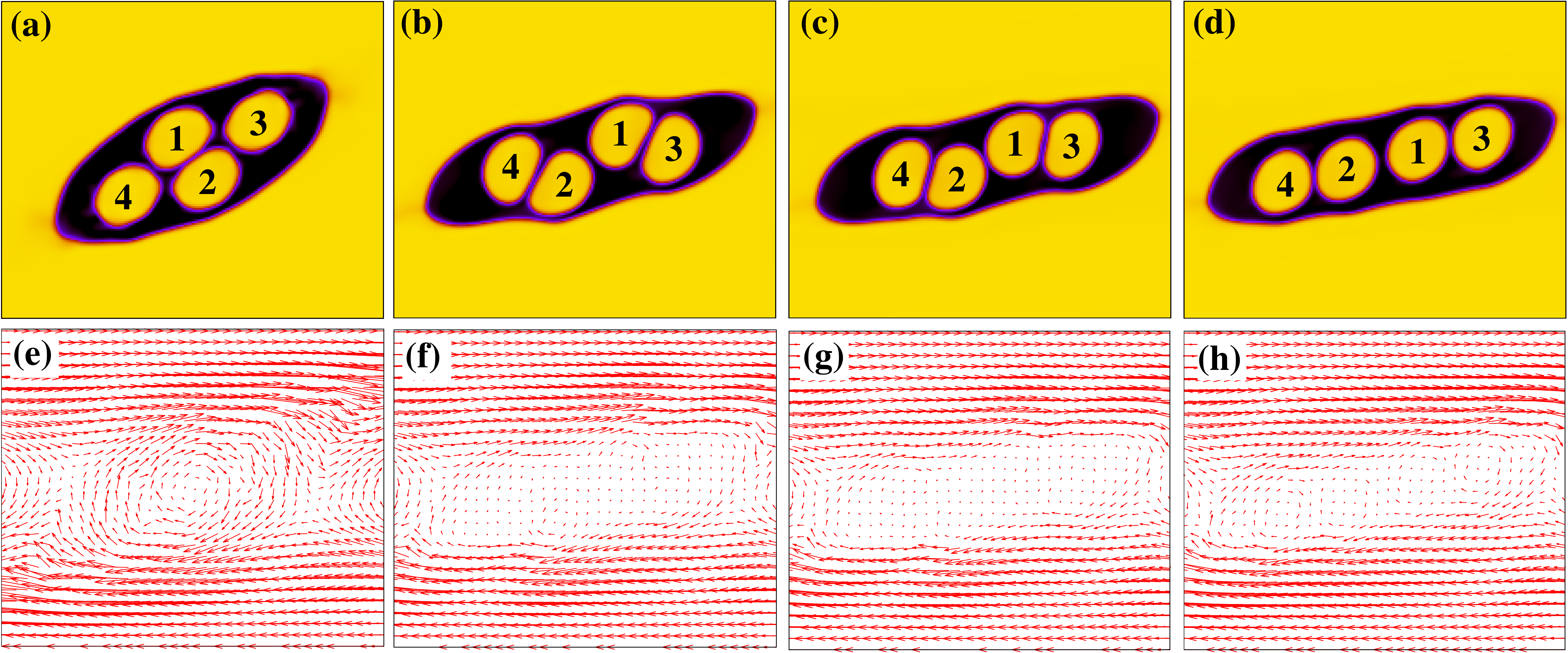}
\caption{Top row: Time evolution of a four-core emulsion under a symmetric shear with $\dot{\gamma}\simeq 2.4\times 10^{-4}$ ($v_w=0.02$). Cores initially rotate clockwise (a), and afterwards gradually align along the shear flow (b-d) due to the large stretching of the external interface which prevents further rotation. Bottom row: Velocity field profile under shear. The internal vortex progressively flattens, and the resulting steady-state pattern resembles that of a single fluid under shear. Snapshots are taken at time (a) $t^*=2$, (b) $t^*=7$, (c) $t^*=19$ and (d) $t^*=31$. Here $Ca\simeq 0.4$ and $Re\simeq 2.4$.}
\label{fig5}
\end{figure*}

In Fig.\ref{fig5} (and \cite{Suppl}, movie M2) we show, for example, the dynamic evolution of a four-core emulsion for $\dot{\gamma}\simeq 2.4\times 10^{-4}$. Here the diffusion constant is $\Gamma=Ma=3.5\times 10^{-4}$ with $M=5\times 10^{-3}$, and it is lower than the value of the previous case. Note that diminishing $\Gamma$, at fixed shear rate $\dot{\gamma}$, increases the Peclet number $Pe=D_Ov_{max}/\Gamma$, a dimensionless number essentially controlling  the mass advection rate with respect to the rate of diffusion. Here, for example, $Pe\simeq 2850$ while it is $\simeq 140$ in Fig.\ref{fig2}. At higher $Pe$ emulsion deformations are expected to be larger, since the mass diffusion rate is considerably lower than the advection rate, despite the system being at moderate values of $\dot{\gamma}$. As long as numerical stability is guaranteed, this ensures that Reynolds and capillary numbers remain well within the laminar regime, without the need of using very high values of $\dot{\gamma}$.

\begin{figure}
\includegraphics[width=1.\linewidth]{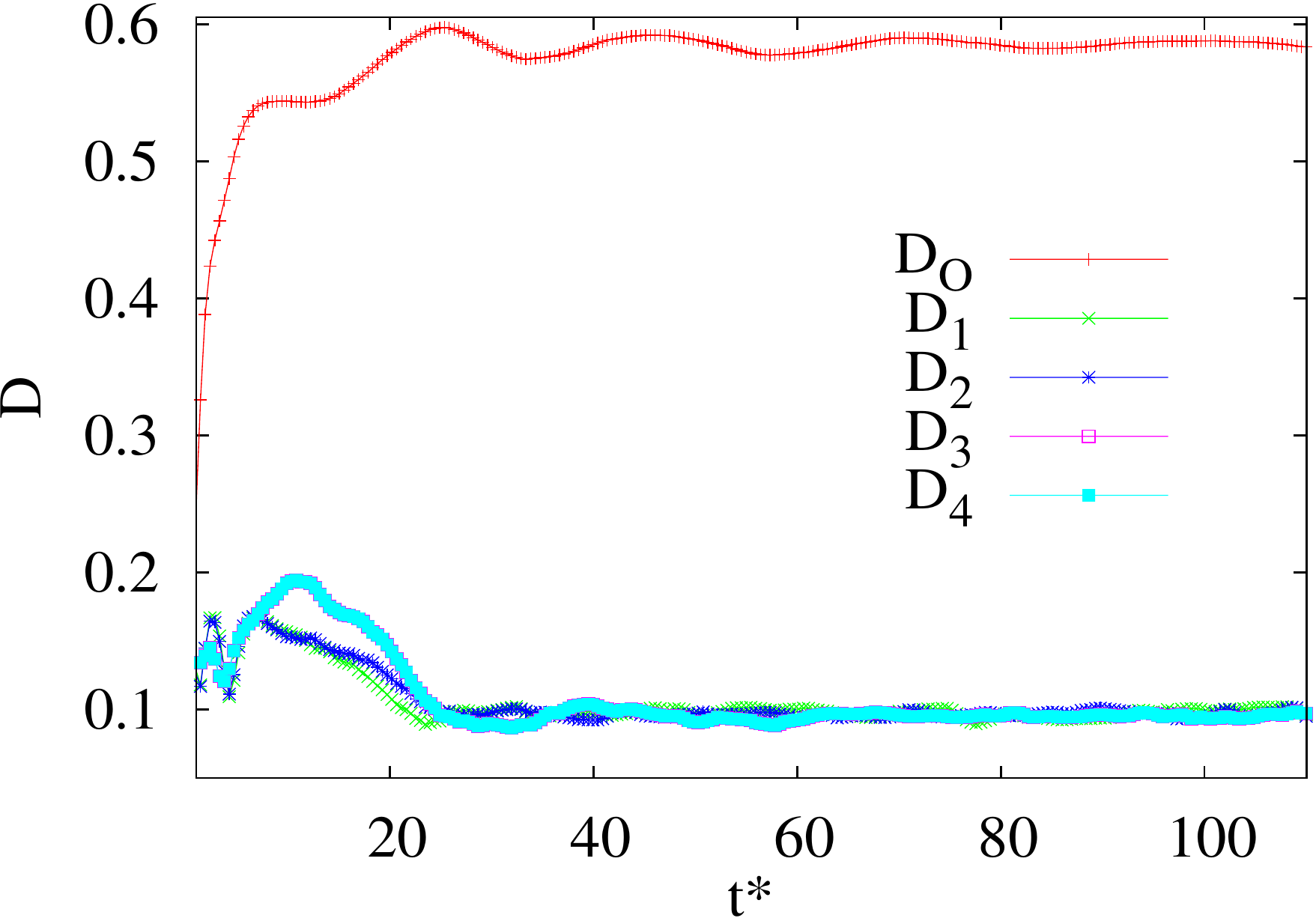}
\caption{Time evolution of the Taylor parameter of the outer droplet $D_O$ and of the cores $D_1$, $D_2$, $D_3$, $D_4$. The former value is approximately six times higher than the latter ones. Time oscillations at early times progressively weaken, and $D$ stabilizes to a roughly constant value.}
 \label{fig6}
\end{figure}

\begin{figure}
\includegraphics[width=1.\linewidth]{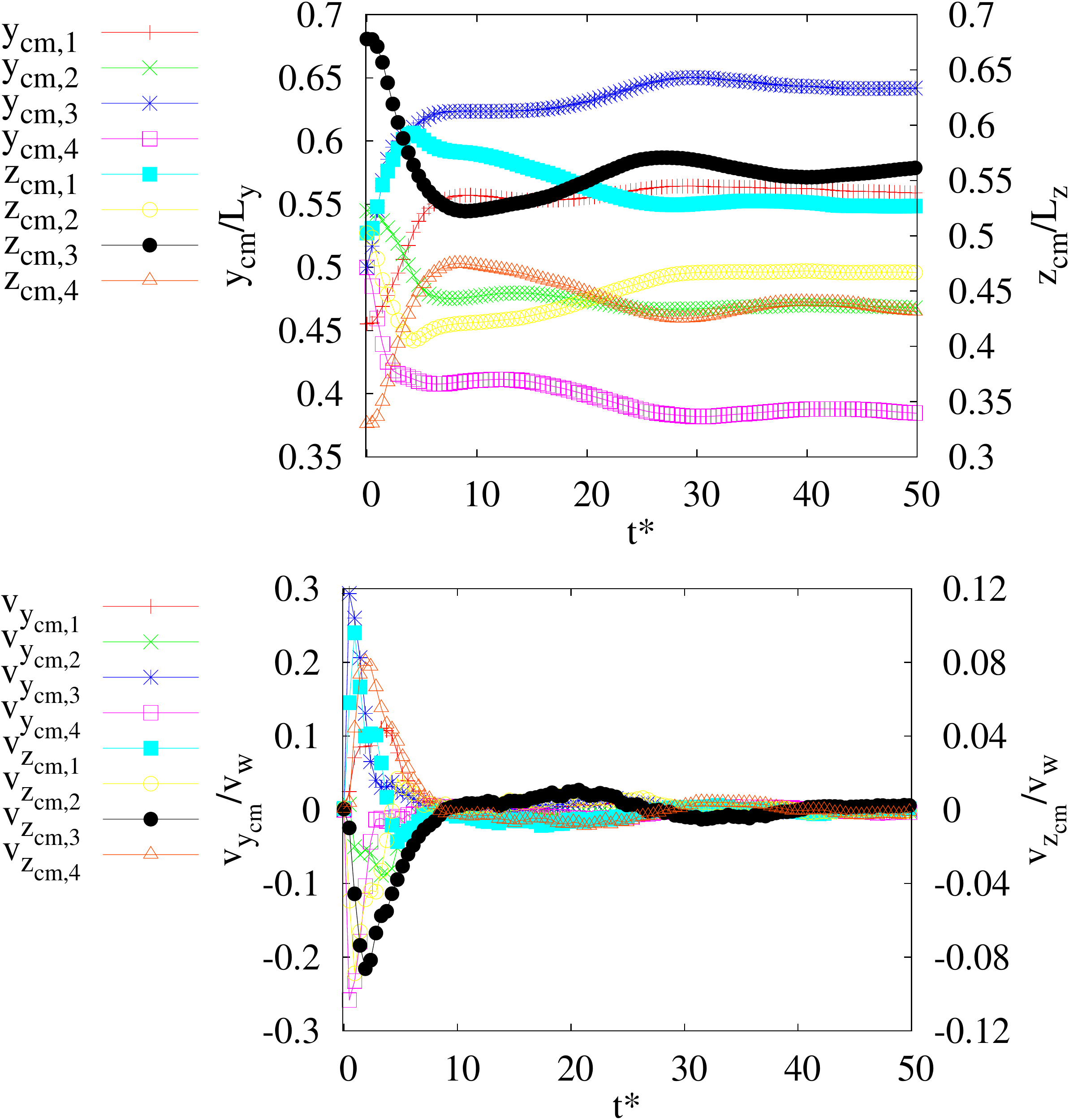}
\caption{Top: Time evolution of the $y$ (left axis) and $z$ (right axis) components of the center of mass of the cores. Oscillations progressively flatten at late times, since cores stabilize and align along the shear direction. Bottom: Time evolution of the $y$ (left axis) and $z$ (right axis) components of the velocity of the center of mass of the cores. A quick increase of the speed, at early times, is followed by its rapid decay to zero at the steady state.}
\label{fig7}
\end{figure}

Like the previous case, once the shear is switched on, the outer droplet tilts along the flow direction and internal cores acquire motion triggered by the fluid vorticity. However, since the flow dramatically increases the eccentricity of the external droplet, such dynamics does not survive at late times. A large deformation of the outer interface (almost six times higher than that of the cores, Fig.\ref{fig6}) squeezes the middle fluid and drastically diminishes the mobility of cores. This effect favours the formation of a quasi-stationary chain-like aggregate of drops aligned along the shear direction and essentially motionless (see Fig.\ref{fig5}a-d and Fig.\ref{fig7}). The fluid recirculation within the emulsion is gradually replaced by an approximately bidirectional flow, reminiscent of the one of a single fluid under shear (see Fig.\ref{fig5}e-h), and the speed of each core rapidly decays to zero, after an initial quick increase caused by the sheared structure of the flow in the channel (Fig.\ref{fig7}, bottom). Note finally that oscillations of $D$ progressively flatten since collisions among internal cores progressively vanish.

The existence of this further steady state suggests that the dynamic response of a multiple emulsion displays a non-trivial dynamics which, even at low/moderate shear, decisively depends on the number of internal drops as well as on long-range hydrodynamic interactions. In the next section we investigate the dynamic behavior of a more complex example of CPE, in which a significantly higher number of cores is included.

\subsection{Higher complex concentrated phase emulsion}

Here we discuss the dynamics under a symmetric shear of the concentrated phase emulsion shown in Fig.\ref{fig1}b, in which nineteen cores, each of diameter $D_i=20$ lattice sites, are included in a larger drop of diameter $D_O=136$ lattice sites. The area fraction is $A_f\simeq 0.4$. Now $\dot{\gamma}$ is varied from $8\times 10^{-5}$ to $2\times 10^{-4}$ ($L_z=250$), thus $Re$ ranges from to $\sim 1.6$ to $\sim 5$, and $Ca$ from $\sim 0.2$ to $\sim 0.62$.

\begin{figure}
\includegraphics[width=1.\linewidth]{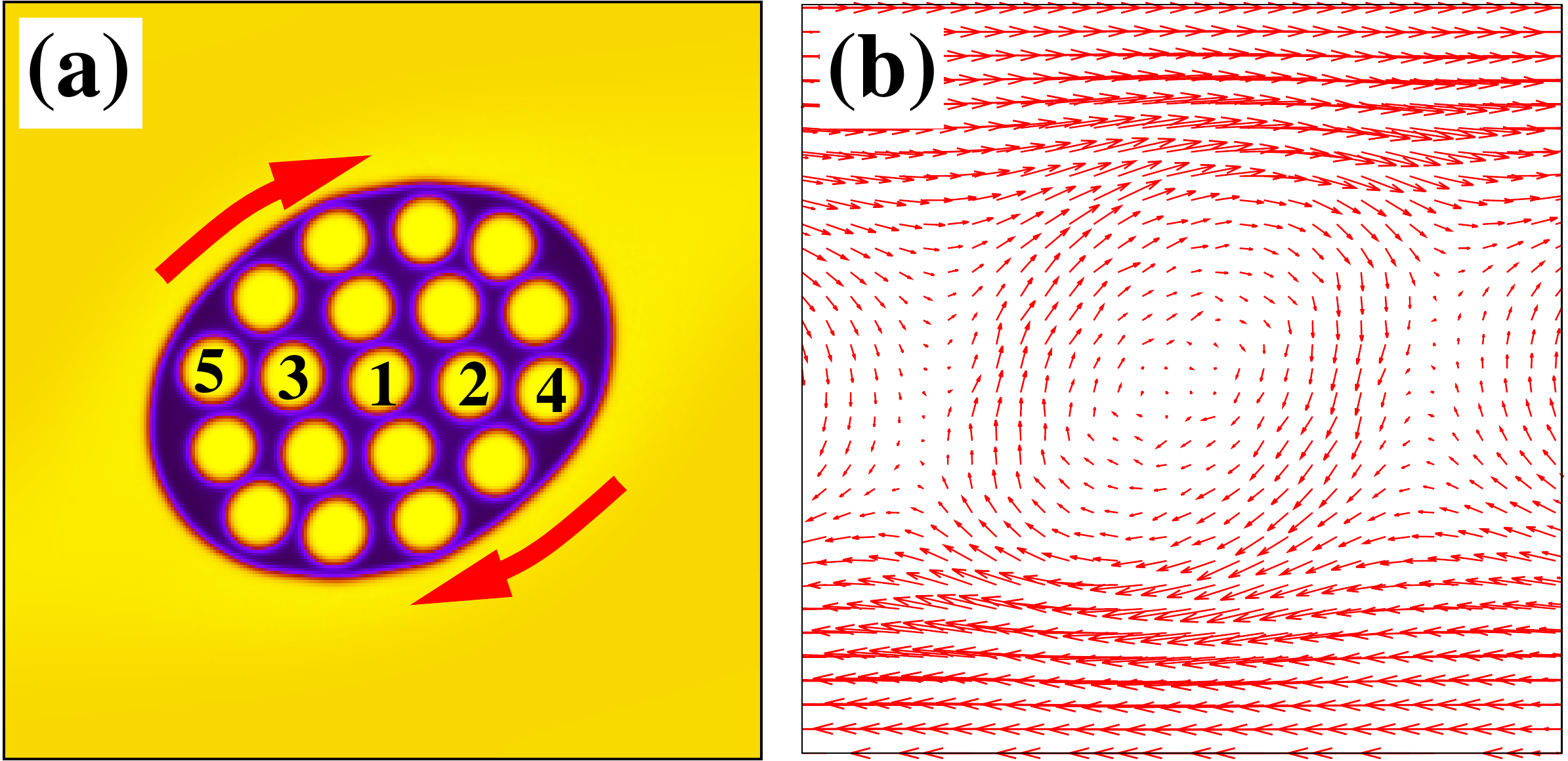}
\caption{(a) Typical nonequilibrium steady-state  of a CPE with nineteen cores under a symmetric shear flow with $\dot{\gamma}\simeq 8 \times 10^{-5}$ ($v_w=0.01$). Once again, a clockwise rotation of the internal cores is triggered by a fluid recirculation produced by the shear within the emulsion. Red arrows indicate the direction of rotation. (b) Steady state velocity field under shear. Snapshots are taken at $t^*=11$. Here $Ca\simeq 0.2$ and $Re\simeq 1.6$.}
 \label{fig8}
\end{figure}

\begin{figure}
\includegraphics[width=1.\linewidth]{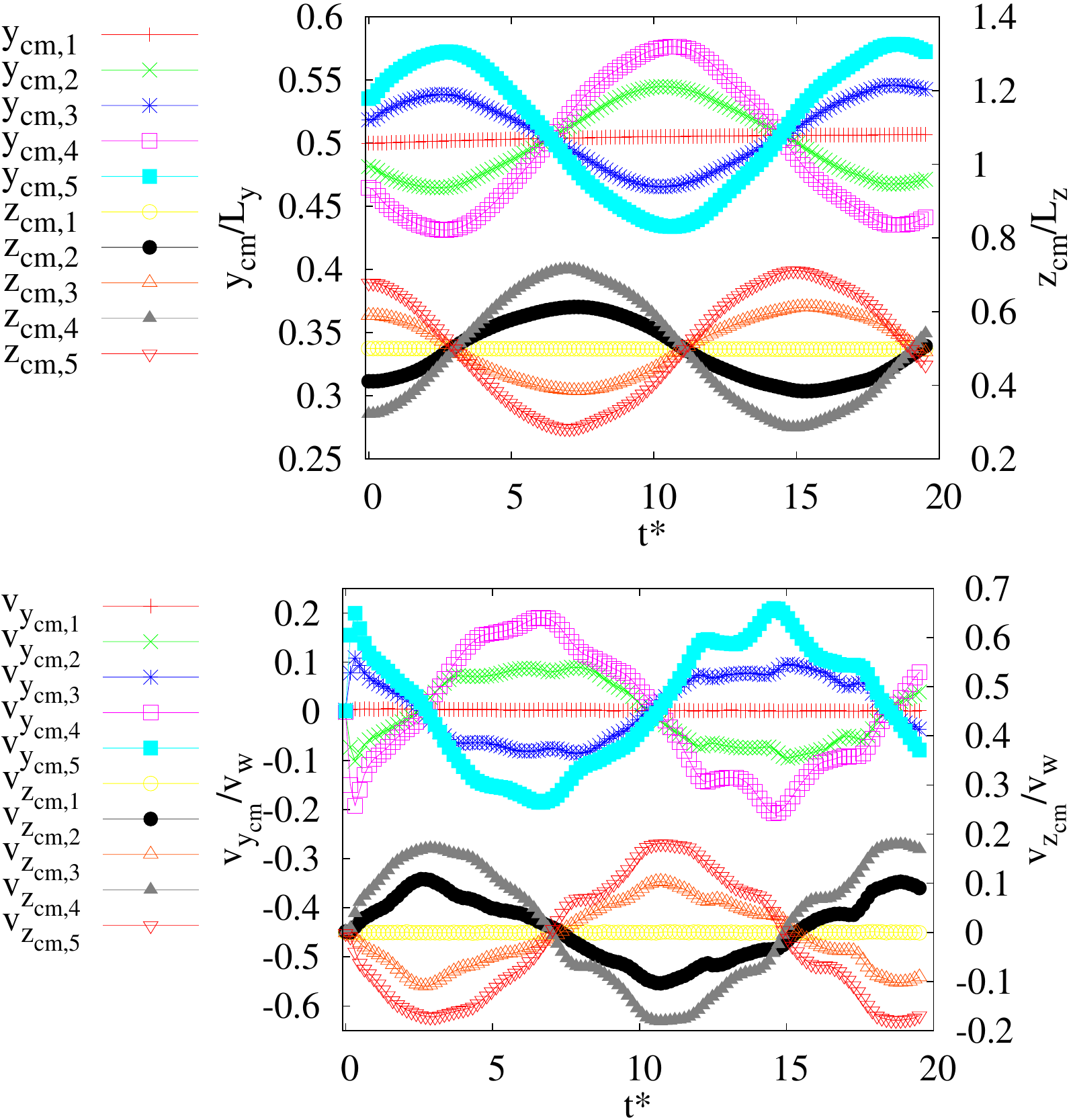}
\caption{Top: Time evolution of the $y$ (left axis) and $z$ (right axis) components of the center of mass of five cores (those reported in Fig.\ref{fig8}). All cores follow a periodic dynamics, but the peripheral ones travel along trajectories larger than the ones covered by inner cores. Bottom: Time evolution of the $y$ (left axis) and $z$ (right axis) components of the velocity of the center of mass. Inner cores move at a speed lower than the one of the peripheral cores. Central core is essentially motionless.}
\label{fig9}
\end{figure}

\begin{figure}
\includegraphics[width=1.\linewidth]{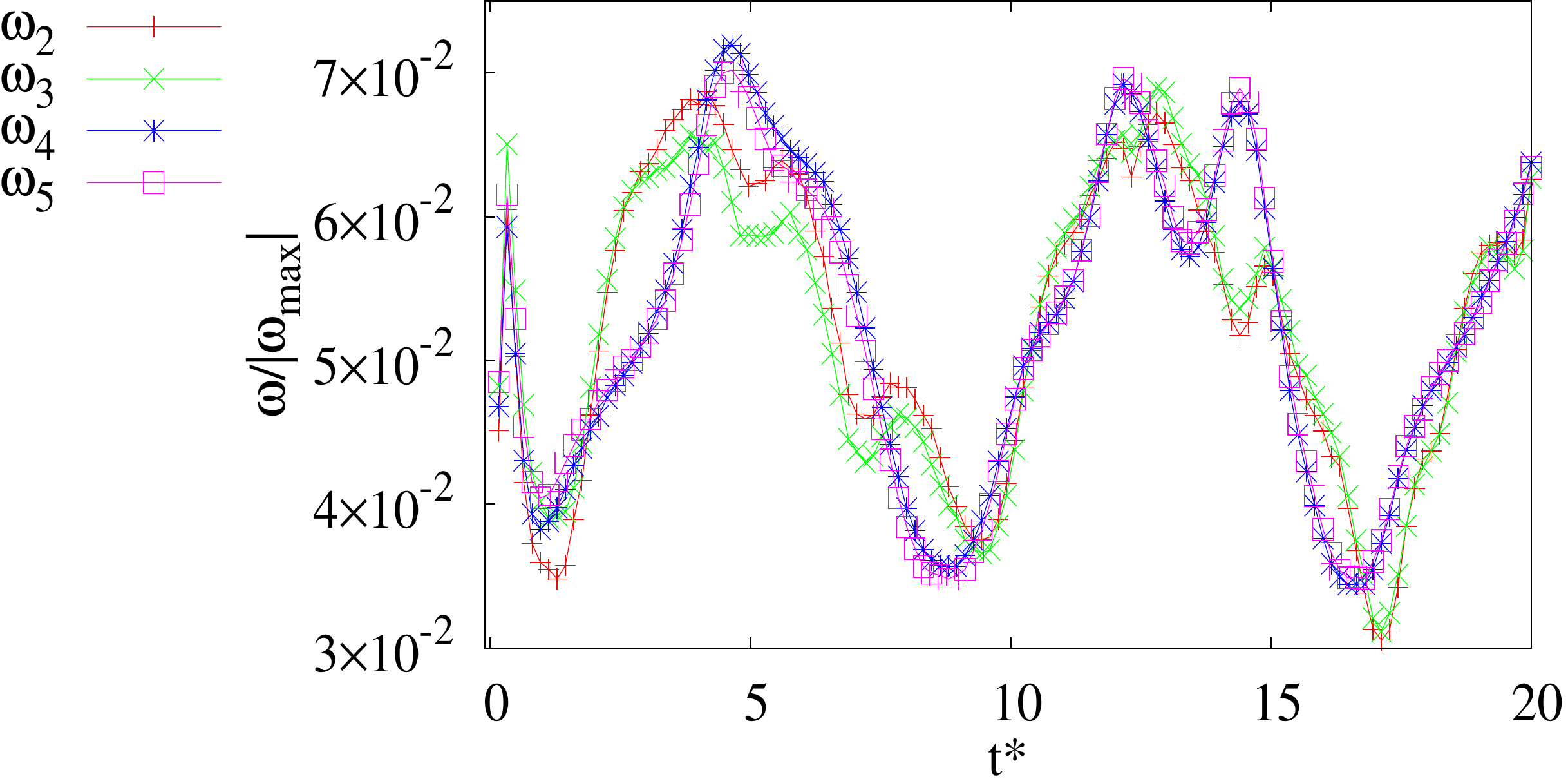}
\caption{Typical time evolution of the angular velocity $\omega$ of four cores of the emulsion in Fig.\ref{fig8}, calculated with respect to $|\omega_{max}|\simeq 0.006$. Peripheral and internal cores travel along different paths with approximately equal values of $\omega$.}
\label{fig13}
\end{figure}

In Fig.\ref{fig8} (and \cite{Suppl}, movie M3 for the complete dynamics) we show a typical nonequilibrium steady state configuration achieved by the nineteen-core emulsion under  shear  with $\dot{\gamma}\simeq 8\times 10^{-5}$. Once again, cores rotate periodically clockwise due to the internal fluid recirculation, but at a speed depending on the position occupied within the emulsion (see Fig.\ref{fig9}). Indeed, peripheral cores (those near to the external interface, such as $4$ and $5$ in Fig.\ref{fig8}a) follow a quasi-circular trajectory larger than the one covered by those (like cores $1$, $2$ and $3$ in Fig.\ref{fig8}a) located close to the center of the emulsion. However, the latter ones travel at a lower speed, and cover a shorter path employing roughly the same time as the outer cores (Fig.\ref{fig9}). In other words, the whole emulsion behaves as a ridig-like body, in which the angular velocity of inner and peripheral cores is approximatel equal. This is shown in Fig.\ref{fig13}, where the time evolution of the angular speed, computed as $\omega_i(t)=v_{cm,i}(t)/(r_{cm,i}(t)-r_{cm,O}(t))$, is calculated for two inner and two peripheral cores.

Such periodic dynamics holds as long as the shear rate is kept sufficiently low. Increasing $\dot{\gamma}$ is expected to produce larger shape deformations, potentially yielding to the breakup of the outer droplet and the release of the cores. However, like the lower-core counterparts \cite{tiribocchi_pof}, before such dramatic event further intermediate nonequilibrium steady states, at low values of $Re$, can be observed.

\begin{figure*}
\includegraphics[width=1.\linewidth]{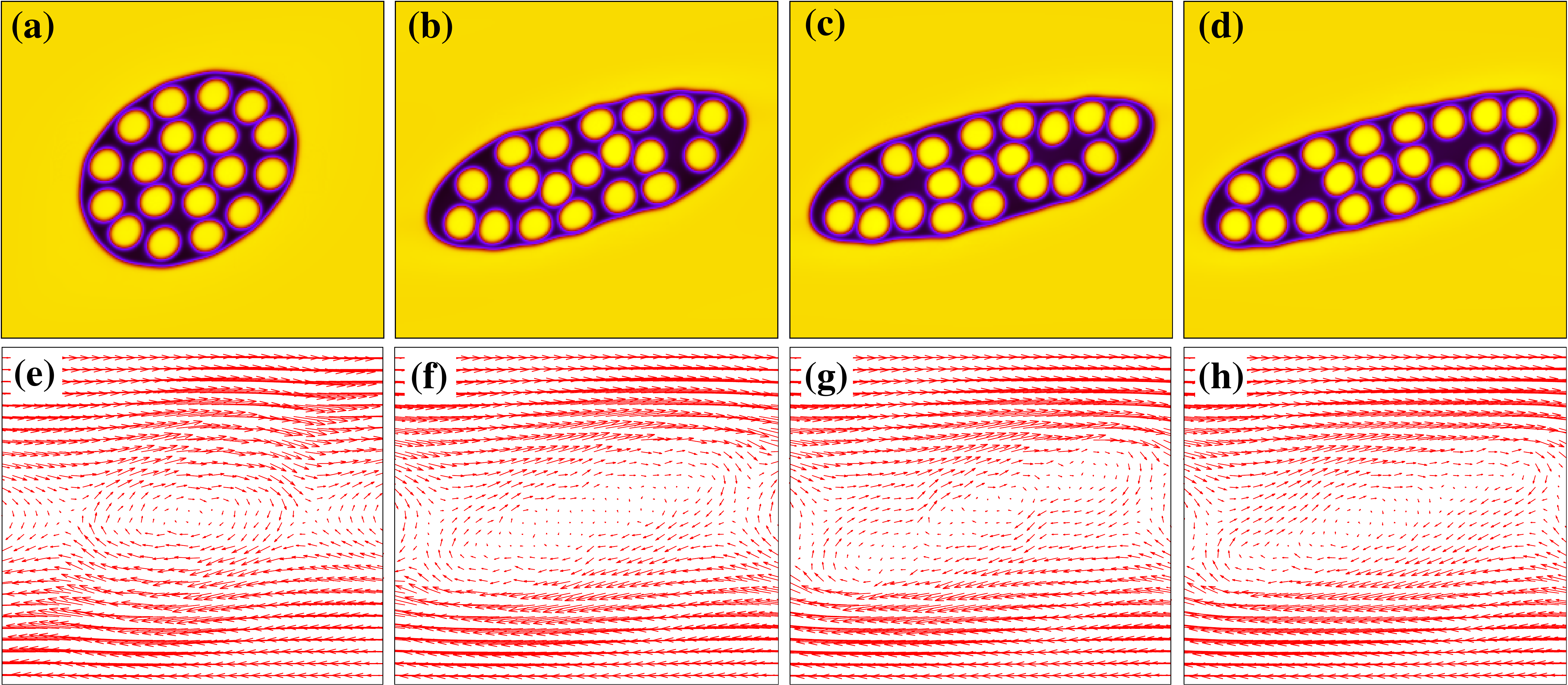}
\caption{Top row: Time evolution of a nineteen-core emulsion under a symmetric shear with $\dot{\gamma}\simeq 2.4\times 10^{-4}$ ($v_w=0.03$). The outer droplet preserves its circular shape only for a short period of time (a), since, due to the shear flow, it gradually streches and attaines, at the steady state, an elliptical-like geometry (b-d). While initially all internal cores rotate clockwise (a), later (b-d) some remain in the center of the emulsion and others accumulate close to the outer interface and rotate clockwise. Bottom row: Velocity field profile under shear. Like in Fig.\ref{fig5}, the internal vortex progressively flattens, although, at the steady state, its structure exhibits weak deviations from a plain elliptical pattern due to the motion of the internal cores which perturb the field. Snapshots are taken at time (a) $t^*=1$, (b) $t^*=8$, (c) $t^*=13$ and (d) $t^*=34$. Here $Ca\simeq 0.6$ and $Re\simeq 5$.}
\label{fig10}
\end{figure*}

\begin{figure*}
\includegraphics[width=1.\linewidth]{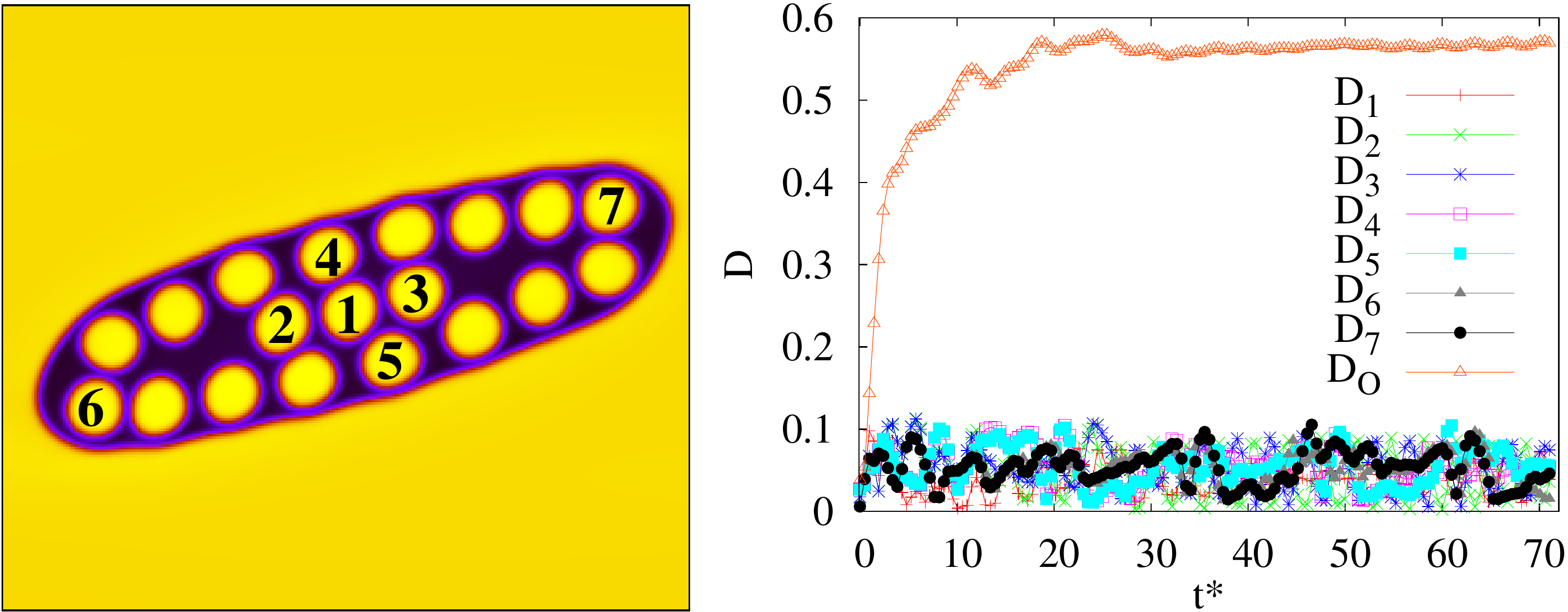}
\caption{Left: Typical steady-state configuration of the multi-core emulsion taken at $t^*=34$. Right: Time evolution of the deformation parameter $D$ of the outer droplet ($D_O$) and of the cores (from $D_1$ to $D_7$). At the steady state, the former is roughly six times higher than the latter.}
\label{fig11}
\end{figure*}

\begin{figure*}
\includegraphics[width=1.\linewidth]{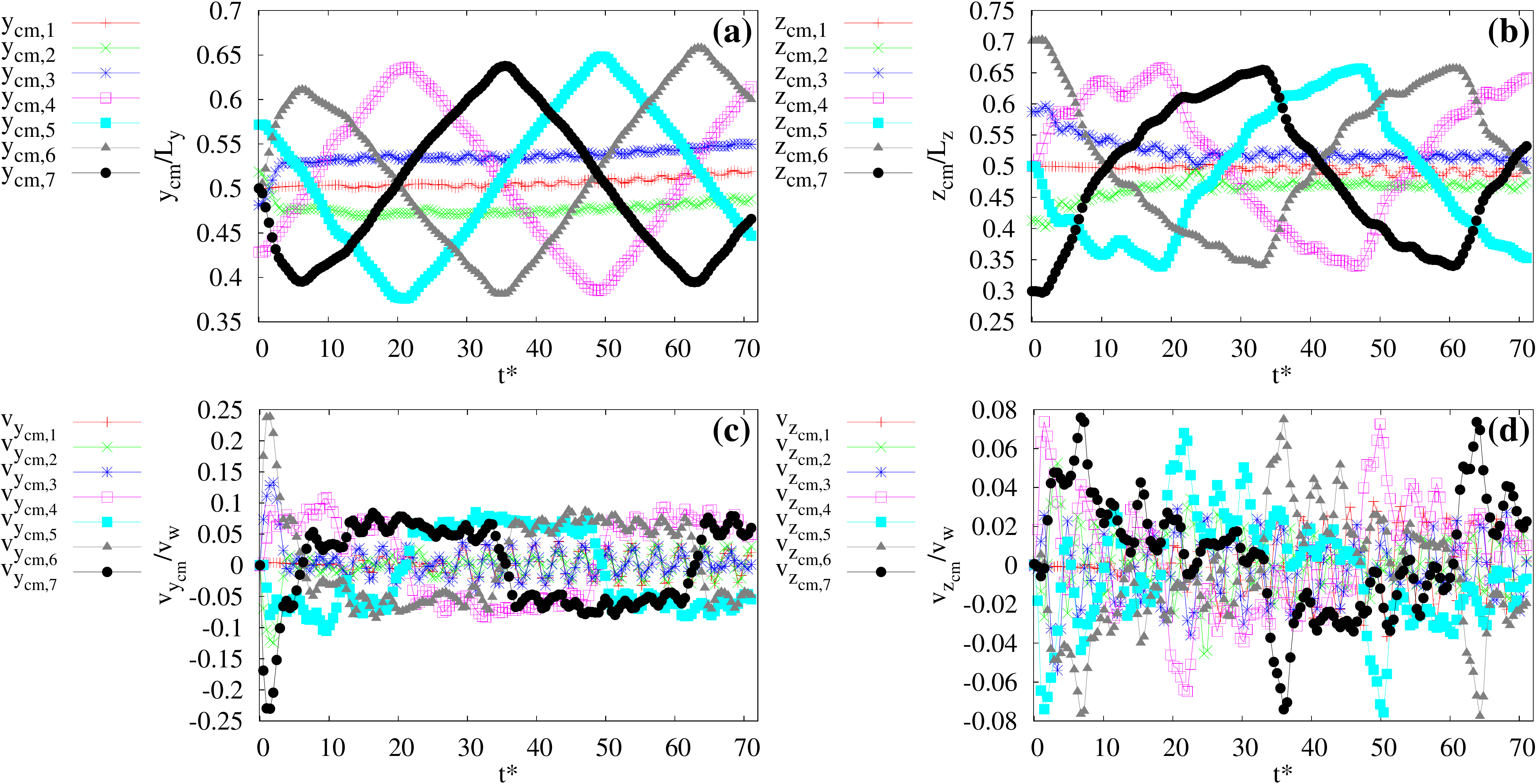}
\caption{Top: Time evolution of the $y$ (a) and $z$ (b) components of the center of mass of the cores numbered in Fig.\ref{fig11}. Bottom. Time evolution of $y$ (c) and $z$ (d) components of their center of mass velocity. At the steady state, three cores very weakly fluctuate in the middle of the emulsion while the others rotate periodically clockwise, with $v_{y_{cm}}$ twice as large as $v_{z_{cm}}$.}
\label{fig12}
\end{figure*}

In Fig.\ref{fig10} (and \cite{Suppl}, movie M4) we show, for example, the dynamic evolution of the emulsion with $\dot{\gamma}\simeq 2.4\times 10^{-4}$. Clearly, here the shear rate is high enough to produce considerable deformations of the outer droplet, which progressively attains an elliptical-like geometry with a value of $D$ approximately six times higher than that of the cores at the steady state (Fig.\ref{fig11}). The solid-like behavior holds for a short period of time, as long as the emulsion remains approximately circular shaped (Fig.\ref{fig10}a). Afterwards, the external interface stretches along the shear flow and the cores, dragged by the fluid current, gradually rearrange. In particular, the peripheral ones accumulate close to the interface and form a dynamic ring chain rotating periodically along elliptical trajectories. On the other hand, the ones located in the middle of the emulsion only temporarily align along the shear direction (Fig.\ref{fig10}b and Fig.\ref{fig12}). Later on, some fill the ``voids'' occasionally left within the chain and get in line, and others remain in the middle of the emulsion due to the lack of space, and are recursively hit by the surrounding drops. At the steady state, the peripheral drops exhibit a periodic dynamics reminiscent of a treadmilling-like motion, while the central ones display weak but regular perturbations caused by the interactions with remaining cores. Finally, unlike the low shear rate case, here the speed of the cores shows irregular and short peaks, a further signature of the recurring interactions among drops and with the external interface.

\subsection{Shapes and dynamic regimes}

\begin{figure*}
\includegraphics[width=1.\linewidth]{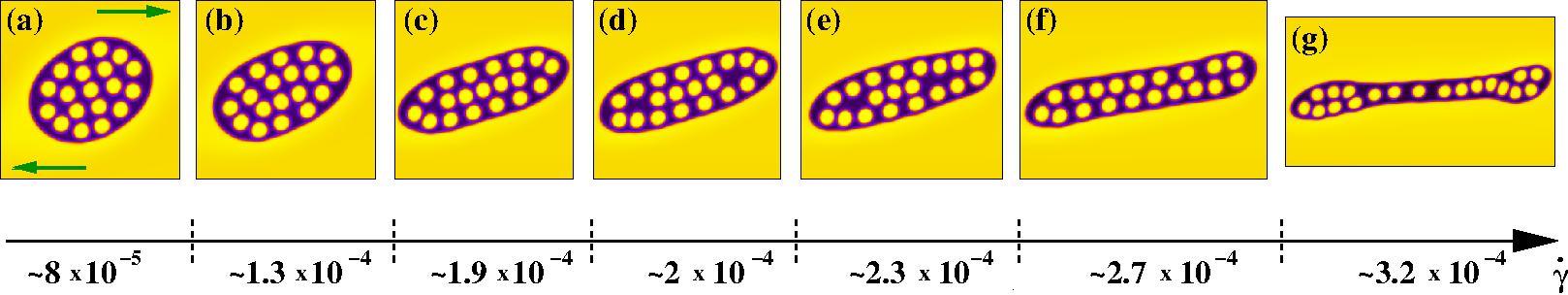}
\caption{Approximate nonequilibrium steady states of a nineteen-core emulsion attained for different values of $\dot{\gamma}$. Green arrows indicate the direction of the moving walls. This applies to all snapshots. For increasing values of the shear rate, the shape of the emulsion varies from a weakly eccentric ellipse (a) towards a mild (b,c,d) and a highly stretched one (e-f), up to a state (g) characterized by two bulges connected by a thinner layer of fluid. The cores generally move along elliptical trajectories but, at high values of shear rates (c,d,e), some accumulate in the middle of the emulsion where are repeatedly hit by the ones in the surroundings. At very high $\dot{\gamma}$, the regular circular motion is replaced by a more chaotic one (g), in which some cores aggregate and rotate within the bulges of the stretched droplet while others fluctuate in the connecting layer of fluid.}
\label{fig14}
\end{figure*}

The nonequilibrium states described so far have provided two typical examples of configurations attained by the emulsion under low and moderate values of $\dot{\gamma}$. A more detailed overview of shapes and dynamic regimes observed by varying $\dot{\gamma}$ is reported in Fig.\ref{fig14}.

As mentioned above, the solid-like behavior described in Fig.\ref{fig8} holds as long as $\dot{\gamma}$ is approximately equal to $1.2\times 10^{-4}$ (Fig.\ref{fig14}a). For $1.3\times 10^{-4}\le\dot{\gamma}\leq 1.5\times 10^{-4}$, the central drop, previously at rest, is set in motion due to the many-body collisions with surrouding cores, and the emulsion finally attains a nonequilibrium state in which all drops periodically travel clockwise along two different elliptical trajectories (Fig.\ref{fig14}b). For intermediate values of $\dot{\gamma}$ (ranging from $\sim 2\times 10^{-4}$ to $\sim 2.5\times 10^{-4}$, Fig.\ref{fig14}c,d,e), the shear flow further stretches the emulsion which, at late times, exhibits a dynamic behavior similar to the one described in Fig.\ref{fig10}. Basically, some cores periodically rotate and remain in the vicinity of the external interface while others align in a row in the middle of the emulsion and, as $\dot{\gamma}$ increases, join the former. This occurs since higher values of shear rate favours the elongation of the emulsion, thus augmenting the distance among peripheral cores as well as collisions against middle-line drops, which are pushed towards the external interface and orbit accordingly.

\begin{figure}
\includegraphics[width=1.\linewidth]{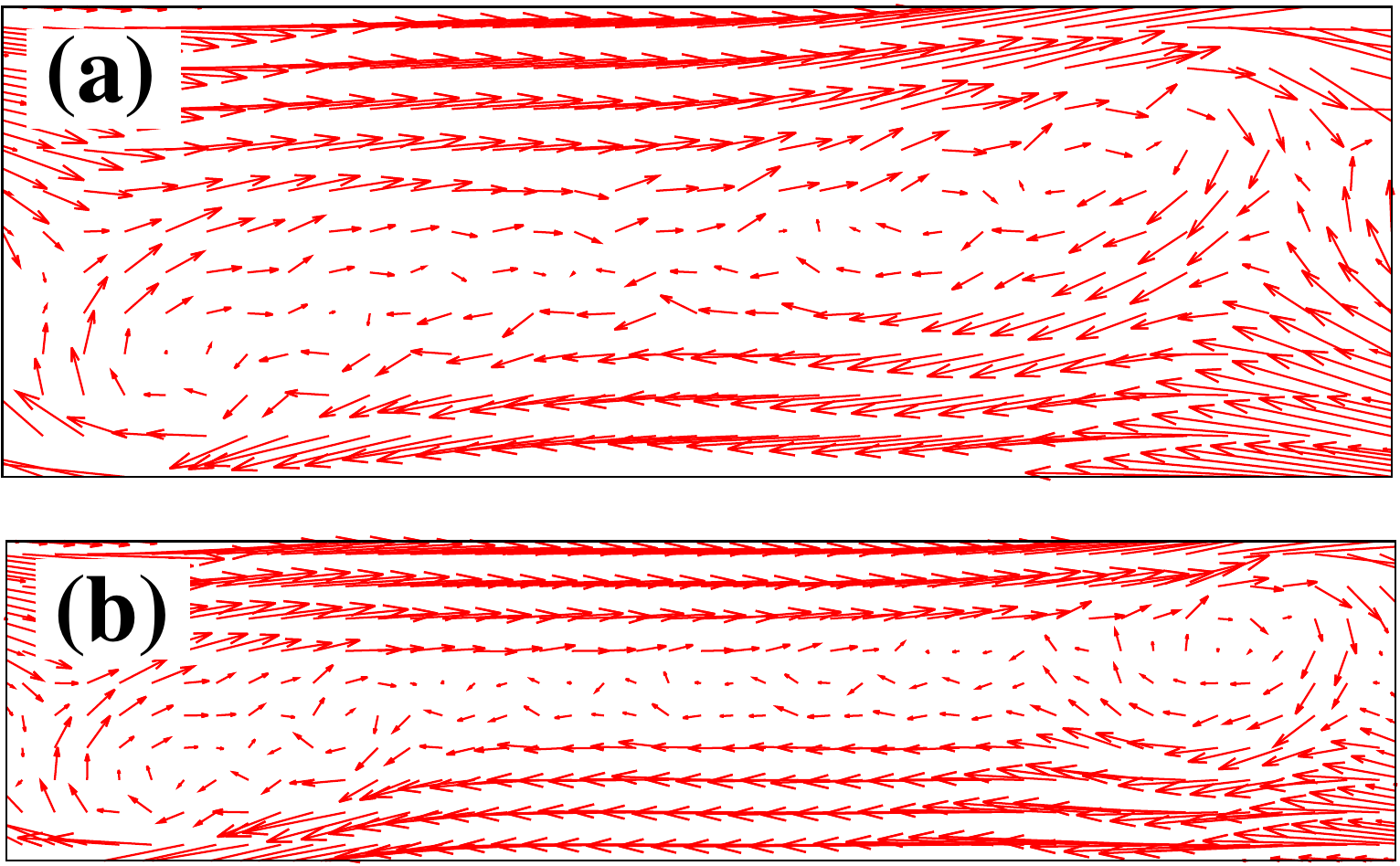}
\caption{Velocity profiles observed within and in the surrounding of the emulsions shown in Fig.\ref{fig14}f and Fig.\ref{fig14}g. The highly stretched recirculation structure shown in (a) is replaced, in (b), by two separate clockwise vortices located at the extremities of the emulsion and connected by a thin layer of fluid where a weaker and irregular fluid pattern  is found.}
\label{fig15}
\end{figure}

At high values of $\dot{\gamma}$ (larger than $\sim 2.5\times 10^{-4}$, with $Re\sim 5.5$ and $Ca\sim 0.7$), the emulsion attains a tank-treading-like configuration (Fig.\ref{fig14}f),  triggered by a highly stretched fluid recirculation pattern consequence of the sheared structure of the fluid (see Fig.\ref{fig15}a and \cite{Suppl}, movie M5, showing the full dynamic behavior). Here the heavy elongation of the emulsion forces a two-row arrangement of the cores which move, once again, clockwise but along a highly-eccentric elliptical trajectory.

The regular dynamics described so far is finally lost by further increasing $\dot{\gamma}$, a regime where inertial effects become relevant. A possible nonequilibrium state attained at late times is that reported in Fig.\ref{fig14}g, where the emulsion has undergone a dramatic extension due to the heavy flow (here $Re\sim 6.5$ and $Ca\sim 0.85$, see \cite{Suppl}, movie M6). A thin layer of fluid links  two separate rounded-shaped bulges located at the extremities (Fig.\ref{fig15}b), in which cores are found to rotate due to the fluid vortices in a periodic manner,  despite the numerous collisions. Occasionally, they are pushed towards the connecting layer where they either weakly fluctuate due to the irregular structure of the velocity field or are sporadically recaptured within the bulges.

These results suggest that the dynamic response of a multi-core emulsion under shear may range among four scenarios. At low shear rate, the emulsion approximately behaves in a solid-like manner since cores move by periodic circular motion at equal angular speed.  For intermediate values of $\dot{\gamma}$, some cores form a dynamic ring chain rotating close to the external interface and others, essentially motionless, align, side by side, along the shear flow. At higher values of shear rate all cores are found to exhibit a treadmilling-like motion occurring along a heavily stretched elliptical trajectory while, for further increasing values of $\dot{\gamma}$, the regular dynamics generally sustained by fluid vortices is replaced by a combination of circular motion and weak fluctuations, observed within far apart rounded bulges connected by a layer of  fluid.

\section{Conclusions}

In conclusion, we have numerically studied the dynamics of a 2D concentrated emulsion with multi-core morphology under a symmetric shear flow in the low/mild Reynolds number regime. Although the shear rate is kept at low/moderate values to avoid the breakup of the emulsion, significant shape deformations occur and new nonequilibrium steady states, previously undetected, are observed. Their formation depends on number of cores and long-range hydrodynamic interactions, controlled by shear rate and mass diffusivity. 

We have considered two examples of concentrated phase emulsions whose design is inspired by compound droplets experimentally realized in microfluidic channels. We observe that, under low values of shear rates, the external droplet weakly stretches and attains an elliptical shape aligned along the flow, while the internal ones acquire motion and rotate  periodically and in a coordinated manner around a common center of mass. In higher complex multi-core emulsions, this effect highlights a solid-like behavior of the system, since all cores travel with equal angular speed. At higher shear rates, the external droplet generally undergoes larger deformations,  which squeeze the middle fluid and significantly affect the motion of the internal drops. In a low-core emulsion, the latter ones have been observed to align along the flow and become essentially motionless, while in high-core systems a more complex dynamics occurs. Some cores form a dynamic ring chain and accumulate near the outer interface where local bulges are produced, while others occupy the center of the emulsion and are recursively hit by the former ones. 
 
Our analysis shows that the dynamics of concentrated phase emulsions with multi-core morphology is significantly more complex than the one observed in more diluted systems, due to non-trivial coupling between fluid velocity and deformations of the fluid interfaces. We hope that our results prove useful to acquire a deeper understanding about the mechanisms governing the dynamic response of confined droplets within capsules flowing within microchannels, of potential interest in the design of soft composite materials built from compound emulsions as well as in drug delivery, where the release of a drug stored within the cores can be significantly affected by viscoelastic properties and shape of the middle fluid \cite{utada_2005,pontrelli2020}.

\section*{Acknowledgements}

A. T., A. M., F. B., M. L. and S. S. acknowledge funding from the European Research Council under the European Union's Horizon 2020 Framework Programme (No. FP/2014-2020) ERC Grant Agreement No.739964 (COPMAT). The authors also warmly acknowledge discussions with  S. Aime, D. Weitz, M. Bogdan and J. Guzowski.

\bibliography{biblio}

\end{document}